\documentstyle[a4,12pt]{article}
\newcommand{\bb}{\begin{eqnarray}}
\newcommand{\ee}{\end{eqnarray}}
\newcommand{\p}{\partial}
\newcommand{\bp}{{\bar \p}}

\newcommand{\bh}{{\bar h}}

\newcommand{\Sg}{\Sigma}

\newcommand{\non}{\nonumber}
\newcommand{\ph}{\varphi}

\newcommand{\Lie}[1]{{\cal L}_{#1}}

\newcommand{\Om}{\Omega}

\newcommand{\La}{\Lambda}
\newcommand{\bOm}{{\bar \Omega}}

\newcommand{\th}{\vartheta}
\begin{document}
\begin{titlepage}
\begin{center}
%\vspace*{.2cm}
\hspace*{10cm} TUM-HEP-243/96\\
\hspace*{10cm} LMU-TPW 96-13\\
\hspace{10cm} hep-th/9604013\\
\hspace{10cm} April 1996\\
\vspace*{1cm}

{\LARGE Non-Abelian Duality for Open Strings\footnote{Work partly supported
by the EC programms SC1-CT91-0729 and SC1-CT92-0789}}\\

\vspace{1.2cm}

{\large S. F\"orste$^{\S}$
\footnote{ E-Mail:Stefan.Foerste@physik.uni-muenchen.de},
A.A. Kehagias$^{\sharp}$
\footnote{E-Mail: kehagias@physik.tu-muenchen.de}
 and
S. Schwager$^{\S}$
 \footnote{E-Mail:Stefan.Schwager@physik.uni-muenchen.de}}\\
\vspace{.8cm}

%\vspace{-.2cm}
${}^{\S}$
{ Sektion Physik\\
%\vspace{-.2cm}
Universit\"at M\"unchen\\
%\vspace{-.2cm}
Theresienstra\ss e 37, 80333 M\"unchen\\
Germany}\\
\vspace{.6cm}

%\vspace{-.2cm}
${}^\sharp${ Physik Department \\
%\vspace{-.2cm}
Technische Universit\"at M\"unchen\\
%\vspace{-.2cm}
D-85748 Garching, Germany}\\
\end{center}
\vspace{1cm}

\begin{center}
{\large Abstract}
\end{center}
\vspace{.3cm}

We examine non-abelian duality transformations  in the open  string case.
After gauging the isometries of the target space and
developing the general formalism, we study in details the duals of
target spaces with SO(N) isometries which, for the  SO(2) case, reduces to the
known abelian T-duals.
We apply the formalism to  electrically and magnetically charged
 4D black hole solutions and, as in the abelian case,
dual coordinates satisfy Dirichlet conditions.

\end{titlepage}
\newpage

\section{Introduction}
According to the duality assertion, a string theory compactified on a manifold
$K$ is equivalent to another theory (possibly the same) compactified on the
dual manifold $\bar{K}$.
Duality symmetries were first realized in the spectrum of the closed
bosonic string \cite{kik}. A bosonic string
 compactified on a circle of radius R is equivalent to the theory obtained if
 the string was compactified on a circle of radius 1/R provided that
momentum and winding modes are interchanged. In the case of closed
superstrings it was proven later that T-duality is not, strictly speaking,   a
symmetry but rather it maps the type IIA theory to type IIB and vice
versa \cite{dine},\cite{dai}.
It was subsequently realized that the same duality
symmetry also exists in the case of  non-trivial backgrounds
with  abelian isometries \cite{bus}--\cite{keh}.
Similarly, in this case although T-duality
is a symmetry of the bosonic string, it is not of the supersymmetric
one and again type IIA theory is mapped to type IIB and vice versa
\cite{keh}.

Duality transformations may also be discussed for target spaces with
non-abelian isometries \cite{fj}--\cite{obers}.
 The  general feature in this case
is that the dual space has in general no isometries
indicating that duality transformations may not necessarily be related
with isometries of the target space but rather may have different
origin. At this point one should also recall the duality-like
symmetries of Calabi-Yau spaces \cite{CY}.

The general procedure is the same
as in the abelian case \cite{bus},\cite{rv}.
 Namely, one starts by gauging the isometries
of the target space in the $\sigma$-model action  and imposing the constraint,
by means of a Lagrange multiplier, that the corresponding field strength
vanishes. Integrating out the Lagrange multiplier one then obtains
the original $\sigma$-model action one  started with. Integrating
out the gauge field, the dual theory is obtained which is again
described by a $\sigma$-model action but with a different, even
topologically, target space. This procedure has been proven to be
equivalent to the first order formulation in the abelian as well as in
the non-abelian case. The difference is that in the former case the
dual and the original theory are equivalent as CFT \cite{rv},
after considering some
global aspects of the procedure, while in the later case they are not
\cite{rg}. It should also be noted,
that there exists the dilaton shift \cite{tra} compensating for the
different ``number" of integrations needed to obtain the dual and the
original theory.

In the open string case now, things are quite different
\cite{dai},\cite{sag},\cite{Hor}. Type I
theory for example compactified on a circle of radius R does not have
a T-duality symmetry. However, one can still map the theory by a
T-duality transformation to another alternative type I' theory in
trying to understand its small R behavior \cite{Hor}.
This type I' theory has
extended dynamical objects, the D-branes \cite{dai},
which are the carriers of
the R-R charges \cite{pol}. These results may also be  extended to the case of
 non-trivial backgrounds \cite{klm},\cite{alva},\cite{otto}.
Here, we will consider  the
general case in which the target has non-abelian isometries.

In the next section and in order to establish notation and
conventions we show  how one may gauge the isometries of the open string.
In sect. 3 we discuss the
non-abelian duality  for open strings in general
non-trivial backgrounds with U(1) charges attached at the ends of the
strings. We also give  explicitly the duality transformations for target
spaces with SO(N) isometry group which we employ in sect. 4 for some 4D
black-hole solutions. In sect. 5 we discuss our results
and we make some concluding remarks.

\section{Gauging the isometries of the  open string}
We will consider here open strings described by the $\sigma$-model
action
\bb
S&=&-\frac{1}{2}
\int_\Sigma d^2\sigma \left(G_{mn}\p_a X^m\p^aX^n +
B_{mn}\epsilon^{ab}\p_aX^m\p_bX^n\right)
-\int_{\p\Sigma} ds A_m\p_sX^m+\nonumber \\
&&\frac{1}{4\pi}\int_\Sigma d^2\sigma\ \sqrt{\gamma}
R^{(2)}\Phi -\frac{1}{2\pi}\int_{\p\Sigma} ds k \hat{\Phi} \, ,
\label{openaction}
\ee
where we have parametrized the boundary of the world sheet $\Sigma$
by $s$ and $\p_s$ is the tangent derivative. The fields $X^m,\,
(m,n=1,\cdots,N)$ are coordinates of the N-dimensional target space M with
metric and antisymmetric fields $G_{mn},\, B_{mn}$, respectively,
 and $A_m$ is a
U(1) gauge field with field strength $F_{mn}$. We have also included a
dilaton $\Phi$ ($\hat{\Phi}$) in the bulk (boundary) of the string
where $R^{(2)}$ is the scalar curvature of $\Sigma$ and $k$ is the
geodesic curvature of the boundary $\p\Sigma$.\footnote{From the string theory
point of view there is no difference in $\Phi$ and $\hat{\Phi}$ since the
dilaton  couples to the Gauss-Bonnet density of the world
sheet.}

We will assume now  that M has some isometries,  generated by a set of
Killing vectors $\xi_I=\xi_I^m\p_m,\, (I,J=1,\cdots,D)$ which form
 the algebra of a
D-dimensional group G with structure constants ${f_{IJ}}^K$, i.e.,
\bb
[\xi_I,\xi_J]={f_{IJ}}^K\xi_K\, .
\ee
Then, under the action of the group, the coordinates $X^m$ transform as
\bb
\delta X^m=\epsilon^I\xi^m_I. \label{dtra}
\ee
In the case of neutral strings (with opposite charges
 attached at their ends), we may write the integral over the boundary as an
 integral in the
bulk by simply replacing $B_{mn}$ by $B_{mn}\!-\!F_{mn}$. In this case,
 the action is invariant under (\ref{dtra}) if
\bb
\Lie{{\xi_I}}G_{mn}&=&0 \nonumber \\
\Lie{\xi_I}(B_{mn}-F_{mn})&=&0 \nonumber \\
\Lie{\xi_I}\Phi&=&0 \nonumber \\
\Lie{\xi_I}\hat{\Phi}&=&0 \, . \label{lieg}
\ee
Recalling that the antisymmetric and the U(1) fields are
defined only up to exact two- and one-forms, respectively, we get
\bb
\Lie{\xi_I}B_{mn}&=&\partial_m\omega_{In}-
\partial_n\omega_{Im} \, , \nonumber \\
\Lie{\xi_I}A_m&=&\partial_m\phi_I+\omega_{Im} \,, \label{Lie}
\ee
for some zero- and one-forms $\phi^I$, $\omega_I$, respectively
 \cite{callann}. The rhs of (\ref{Lie}) is invariant under the following
transformations
\begin{equation} \label{residual}
\begin{array}{lll}
\omega_{Im} &\rightarrow  \omega_{Im}^{\prime } & = \omega_{Im} +
\partial_{m}h_I \\
\phi_I &\rightarrow \phi_I^{\prime} & =\phi_I  -h_I + k_I,
\end{array}\end{equation}
where the $k_I$ are constants and the $h_I$ are scalars.
By evaluating $\Lie{[\xi_I,\xi_J]}B_{\mu\nu}$ and
$\Lie{[\xi_I,\xi_J]}A_m$ we find that
 $\omega_I$ and $\phi_I$ must satisfy the
consistency conditions
\bb
\Lie{\xi_I}\omega_J-\Lie{\xi_J}\omega_I&=&{f_{IJ}}^K\omega_K+ v_{IJ}\, ,
\nonumber \\
\Lie{\xi_I}\phi_J- \Lie{\xi_J} \phi_I&=&{f_{IJ}}^K\phi_K-\rho_{IJ}\,
,\label{consistency}
\ee
where $v_{IJ}$ is a closed one-form, i.e.\ it can locally be expressed as
$$ v_{IJ} = d\rho_{IJ}. $$
We know that for an anomaly free gauged model in the closed
string case to exist one should be able to remove $v_{IJ}$ from
(\ref{consistency}) \cite{jack}. This  implies
certain integrability conditions on the $v_{IJ}$.  Assuming that these
conditions are satisfied
we perform a transformation (\ref{residual}) to remove $v_{IJ}$ from the first
equation
in (\ref{consistency}) and then the consistency conditions read
\bb
\Lie{\xi_I}\omega_J-\Lie{\xi_J}\omega_I&=&{f_{IJ}}^K\omega_K\, , \nonumber \\
\Lie{\xi_I}\phi_J- \Lie{\xi_J} \phi_I&=&{f_{IJ}}^K\phi_K-k_{IJ}\,
,\label{consistency1}
\ee
with $k_{IJ}$ being constants. Now, the residual symmetries of (\ref{residual})
are just constant shifts in $\phi_I$ under which the $k_{IJ}$ will transform as
\bb  \label{residual1}
                                                      k_{IJ} \rightarrow k_{IJ}
+ {f_{IJ}}^K k_K \, .
\ee
As we will see later on the constants $k_{IJ}$ have to
satisfy certain integrability conditions.

Let us now gauge the isometries by assuming that the group parameters
$\epsilon^I$ in eq.(\ref{dtra}) are local.
Then, following \cite{jack},
we introduce gauge fields $\Omega_a^I$ on the world sheet and
we demand invariance of the
action under
\bb
\delta X^m&=&\epsilon^I\xi^m_I \, , \nonumber \\
\delta \Omega_a^I&=& \p_a \epsilon^I+{f_{KJ}}^I \Omega^K_a \epsilon^J \, .
\label{gtrans}
\ee
We define covariant  derivatives  by
\bb
D_aX^m = \p_a X^m -\xi^m_I\Om_a^I \, , \label{der}
\ee
such that $D_aX^m$ transforms as
\bb
\delta (D_aX^m)=\epsilon^I\p_n\xi^m_ID_aX^n \, . \label{DX}
\ee
The method to construct a gauge invariant action in the closed string case
is discussed in \cite{jack}. For general $B$ fields the gauge invariant action
is not given
by replacing partial derivatives with covariant ones.
Since here we are mainly interested in modifications due to
the boundary we put from now on the $B$ field to zero and hence minimal
coupling works
for the bulk part of the action,
\bb
S_{bulk} =-\frac{1}{2}
\int_\Sigma d^2\sigma G_{mn}D_a X^mD^aX^n \, ,  \label{gauged}
\ee
where dilaton contributions have been omitted.

By using  the identity $\Lie{\xi}=
\imath_\xi d+d\imath_\xi$ where $\imath_\xi$ is
the inner product in the exterior algebra, we get that the variation
of the action
(\ref{openaction}) under the global transformations eq.(\ref{dtra}) is
\bb
\delta S=-\int d^2\sigma \epsilon^I
\Lie{\xi_I}G_{mn}\,\p_a X^m\p^aX^n
-\int ds\, \epsilon^I\xi^m_I F_{mn}\,\p_sX^n\, .  \label{ds}
\ee
Thus,  (\ref{openaction}) is invariant if
$\Lie{\xi_I}G_{mn}=0$ and if,  in addition,
there exists an exact one-form $dv_I$ such
that
\bb
\xi^m_I F_{mn}=\p_nv_I \, .\label{fff}
\ee
Eq.(\ref{fff}) implies  that $\Lie{\xi_I}F_{mn}=0$ so that
\bb
\Lie{\xi_I}A_m&=&\partial_m\phi_I \,, \label{Liee}
\ee
in agreement with eq.(\ref{Lie}). By using eqs.(\ref{fff},\ref{Liee}), we get
\bb
\xi_I^mF_{mn}&=&\xi^m_I\p_mA_n-\xi_I^m\p_nA_m=-\p_n\phi_I-A_m\p_n\xi^m_I
-\xi_I^m\p_nA_m\nonumber \\
&=& \p_n(\phi_I-A_m\xi^I_m)=\p_nv_I\, ,\label{ses}
\ee
so that $ \phi_I$ and $v_I$ are not independent but, up to a constant,
\bb
A_m\xi^m_I-\phi_I=v_I \, . \label{uh}
\ee

In order to find the gauge invariant action we may
follow Noether's procedure, i.e.,  we first calculate the variation of the
ungauged action and then by adding terms linear  in the gauge field
we try to cancel these variations. If
$S^{(0)}=\int ds A_m\p_sX^m$,
\bb
\delta S^{(0)}=\int_{\p\Sigma} ds\,\left(\epsilon^I
\Lie{\xi_I}A_m\, \p_sX^m+A_m\xi^m_I\p_k\epsilon^I
\p_sX^k \right)
\ee
and  one may  try
 to cancel $\delta S^{(0)}$  by adding the term
$S^{(1)}=\int ds\, C_I\, \Om^I_s$ to $S^{(0)}$ where $\Omega_s$ is the
component of the gauge field in the tangent direction. Then
\bb
\delta S^{(1)}=\int_{\p\Sigma}  ds\,
\left( \p_mC_I\,\Om^I_s\epsilon^J\xi_J^m+C_I\,\p_n\epsilon^I\,\p_sX^n
+C_I{f_{JK}}^I\Om_s^J\epsilon^K\right)\, ,
\ee
and thus,
\begin{equation} \begin{array} {l l}
\delta S^{(0)}+\delta S^{(1)}=&\int_{\p\Sigma} ds
 \left(\epsilon^I\p_m\phi_I\p_sX^m+A_m\xi^m_I\p_k\epsilon^I
\p_sX^k + \epsilon^I\xi_I^m\p_mC_J\,\Om^J_s+ \right. \\
 &\left. C_I\p_m\epsilon^I\,\p_sX^m
+C_I{f_{JK}}^I\Om_s^J\epsilon^K \right) \, .\label{delt}
\end{array} \end{equation}
The condition for gauge invariance turns then out to be
\bb
C_I&=&-A_m\xi^m_I+\phi_I \label{c1}+\lambda_I\\
\Lie{\xi_I}C_J&=&-{f_{JI}}^KC_K\, , \label{c2}
\ee
where $\lambda_I$ are constants.
It is not difficult to verify that (\ref{c2}) is compatible with (\ref{c1}) if
the following condition is satisfied
\begin{equation} \label{kkk}
k_{IJ} -{f_{IJ}}^{K}\lambda_K =0\, ,
\end{equation}
where $k_{IJ}$ is defined in (\ref{consistency1}). The integrability
condition of (\ref{kkk}) follows from the Jacobi identities of the
isometry algebra and reads
\bb
{f_{IJ}}^Kk_{KL} + cycl.\, perm.\, = 0.
\ee
The $\lambda_K$ are not uniquely fixed but can be shifted according to
(\ref{residual1}). As we will see later the freedom of shifting $\lambda_K$
represents the freedom in the position of the D-brane in the dual model.
Thus, by using eq.(\ref{uh}),  the gauge invariant action is
\bb
S=-\frac{1}{2}\int_\Sigma d^2\sigma G_{mn}D_a X^mD^aX^n -
\int ds\, \left(A_m\p_sX^m-v_I\Omega^I_s\right) \, . \label{ginv}
\ee
It should be noted that in the case of unoriented open strings (where B=0)
the gauged action is again given by eq.(\ref{ginv}). It is also obvious that
the gauged action still has the U(1) symmetry $A_m\to A_m +\p_m\ph$ of the
original ungauged one.

\section{Non-abelian T-duality}

Below we examine T-duality transformations for open strings. For this,
we will consider  a $\sigma$-model with vanishing antisymmetric tensor since
its presence  does not modify  the discussion and moreover it is absent in the
examples we will consider later on. Then, in
 null--coordinates $z=\frac{1}{\sqrt{2}}(\tau+\sigma),
\bar{z}=\frac{1}{\sqrt{2}}(\tau-
\sigma)$ ($\p=\p/\p z,\bp=\p/\p\bar{z}$), the $\sigma$-model action is
\bb
S=\int_\Sigma d^2z\,G_{mn}\p X^m\bp X^n
-\int_{\p\Sg} ds A_m \frac{dX^m}{ds} \, , \label{actt}
\ee
where  the dilaton terms have been omitted.
We may  enlarge the target space by adding  fields
$X^\alpha,\, (\alpha,\beta=1,\cdots,d)$
which parametrize a space K so that the
target space is locally  $M\!\times\!K$.
Following the procedure of the previous chapter, one may gauge the isometries
of M (or some subgroup) and then the   gauged action turns out to be
\bb
S(X,\Om)&=&\int_\Sg d^2z\, \left(G_{\alpha\beta}\p X^\alpha \bp X^\beta
+G_{\alpha m}\p X^\alpha \bar{D}X^m+
G_{m\alpha}DX^m \bp X^\alpha+G_{mn}DX^m \bar{D}X^n\right)
\nonumber \\ && + \int_\Sigma d^2z\, \La_IF^I-
\int_{\p\Sg} ds\left( A_\alpha \p_sX^\alpha+
A_m \p_sX^m-v_I(\Omega^I\frac{dz}{ds}+\bar{\Omega}^I\frac{d\bar{z}}{ds})
\right) \, . \label{ga}
\ee
We have also included above the term $\La_IF^I$ where $\La_I$ is a Lagrange
multiplier and  $F=\p \bar{\Om}-\bp \Om+[\Om,\bar{\Om}]$ is the field strength
of the gauge field $\Omega$. It is then straightforward to verify that
(\ref{ga}) is invariant under the local transformations
\bb
\delta X^m&=&\epsilon^I\xi^m_I \, , \nonumber \\
\delta \Omega^I&=& \p \epsilon^I+{f_{JK}}^I \Omega^J \epsilon^K
\, , \nonumber \\
\delta \bar{\Omega}^I&=& \bar{\p} \epsilon^I+{f_{JK}}^I
\bar{\Omega}^J \epsilon^K \, ,
\nonumber \\
\delta \La_I&=&-{f_{IJ}}^K\La_K\epsilon^J \, . \label{dddd}
\ee

The  path integral now for the open string is as usual an integral  over
all field configurations which, however, are
consistent with some specified boundary conditions
\bb
Z(\{S_B\})=\int_{\{S_B\}}DX e^{iS(X)} \label{PI}.
\ee
and it is  a functional of the boundary conditions, collectively
written as $\{S_B\}$. The total vacuum
to vacuum amplitude is then given by an
integration of the $Z(\{S_B\})$'s over
all allowed boundary conditions.
The path integral after gauging is
\bb
\tilde{Z}(\{S_B\})=\int_{\{S_B\}}DX\,\frac{D\Om\,D\bar{\Om}}
{{\cal{V}}}D\La\;e^{iS(X,\Om)}
\ee
where we have divided out the ``volume" ${\cal{V}}$ of the gauge group.
Integration over $\La$ constrains the gauge fields to
be flat, $\Om = h^{-1}dh$ for some $h \in G$. Fixing the gauge by
choosing a section $h(z,\bar{z})$ gives back the ungauged action $S(X')$ with
transformed coordinates $X'=hX$. However, the boundary conditions may have
changed. The requirement that a pure gauge does not modify  them
 will give boundary conditions for  the gauge parameters which can be
implemented by a second Lagrange multiplier. For Neumann  conditions,
for example, one has to restrict the normal
component  of the gauge field  to
vanish  at the boundary. We will not do that here since the final
result does not depend on this second Lagrange multiplier as one can
check by an explicit calculation.

Our next task is to integrate over the gauge fields $\Om$.
For this we perform a
partial integration in the term $\int d^2z\, \La_I F^I$
%\bb
%\int Tr(\La d\Om) &=& \int d\,Tr(\La\Om) - \int Tr(d\La\we\Om) = \non\\
%&=&\oint Tr(\La\Om) - \int Tr(d\La\we\Om).
%\ee
%We assume the group generators normalized as $Tr(T_\al
%T_\beta)=\delta_{\al\beta}$, so we write expressions like $Tr(\La F)$ as
%$\La_\al F^\al$ etc. (Another normalization could be compensated by a
%%rescaling
%of the Lagrangemultiplier $\La$).
and after some algebra, the complete action  is found to be
\bb
S_g(X,\Omega)&=&\int_\Sg d^2\sigma\left(G_{\alpha\beta}\p X^\alpha \bp X^\beta
+G_{\alpha m}\p X^\alpha\bp X^m+G_{m\alpha}\p X^m\bp X^\alpha +
\right. \nonumber \\
&&G_{mn}\p X^m \bp X^n -\left.\Om^I\bh_I -\bOm^I h_I+
\Om^I f_{IJ}\bOm^J \right)
-\int_{\p\Sg}ds\left(A_\alpha\frac{dX^\alpha}{ds} +\right. \nonumber \\
&&\left. A_m\p_sX^m-v_I(\Omega^I\frac{dz}{ds}+\bar{\Omega}^I
\frac{d\bar{z}}{ds})-\La_I\Omega^I\frac{dz}{ds}-
\La_I\bar{\Omega}^I\frac{d\bar{z}}{ds}\right)\, , \label{ca}
\ee
where $h_I,\bh_I$ and $f_{IJ}$ are given by
\bb
h_I &=& \p\La_I+\p X^m G_{mn}\xi_I^n+\p X^\alpha G_{\alpha m}
\xi_I^m\, , \non\\
\bh_I &=& -\bp\La_I+\bp X^m G_{nm}\xi_I^n+\bp X^\alpha
G_{m\alpha}\xi_I^m \, ,\non\\
f_{IJ} &=& {f_{IJ}}^K \La_K +
G_{mn}\xi_I^m\xi_J^n. \label{hh}
\ee
As a result, the path integral is
\bb
Z=\int DX\,\frac{D\Om\,D\bar{\Om}}
{{\cal{V}}}D\La&&\hspace{-.2cm}
\exp\, i\left[S+\int_\Sigma d^2z\left(-\Om^I\bh_I -\bOm^I h_I+
\Om^I f_{IJ}\bOm^J\right)-\right.
\nonumber \\ &&  \left.\int_{\p \Sigma} ds
\left( (\Omega^I\frac{dz}{ds}
+\bar{\Omega}^I\frac{d\bar{z}}{ds})
(\La_I+v_I)\right)\right]\, , \label{part}
\ee
where $S$ is the ungauged action in eq.(\ref{actt}). In order to get the dual
action we have to integrate out the gauge fields
$\Omega^I,\bar{\Omega}^I$. Here, however, due to the boundary term in the
exponent, one has to integrate the gauge field independently in the bulk and on
the boundary \cite{otto}. The integration in the bulk gives the dual action
\cite{qq}
\bb
\tilde{S}[X,\La]&=&S[X]-\int d^2\sigma h_I(f^{-1})^{IJ}\bar{h}_J
\, ,  \label{da}
\ee where $S[X]$ is the $\La$-independent part. Since the gauge field appears
linearly in the boundary term (the last term in eq.(\ref{part})),
its integration
will produce a delta function \cite{otto},
i.e., the constraints
\bb
\La_I+v_I=0 \, \hspace{.5cm} on\,\,\, \p\Sigma \, .
\label{dirr}
\ee
As a result, the dual fields have to satisfy Dirichlet conditions, no matter
what boundary conditions the original fields satisfy. This means that in the
dual theory the boundary conditions are imposed as a constraint irrespectively
of any stationary conditions of the action.

As follows from eq.(\ref{c2})  $v_I$
transforms in the same representation of the isometry group as the $\La_I$ so
that  the boundary condition eq.(\ref{dirr}) is covariant. Hence,
the position of the  D-brane is  gauge dependent.
 Moreover, as was mentioned before,   $v_I$
is defined only up to a constant  corresponding  to the U(1) symmetry
of the  original model.
In the dual theory  this symmetry  manifests itself
as shifts in the position of the D-brane.

Although it is not possible to give general formulas like in the
abelian case, one has to  discuss  specific examples in order to illustrate the
method.
Below we will give the calculation for the SO(N) case
since we will use it later to recover the abelian case and to find the
duals of 4D black holes in the open string case.
%It should be noted that the bulk dilaton, as in the closed string case
%\cite{qq}, \cite{tra}, changes here as well under this process as
%\bb
%\Phi\rightarrow \Phi^\prime=\Phi-\frac{1}{2}\ln det(f)\, ,\label{dil}
%\ee
%where $det(f)$ is the determinant of the matrix $f_{IJ}$ of eq.(\ref{hh}).
%In is of course expected that the dilaton on the boundary has a similar
%transformation so that the correct coupling to the Gauss-Bonnet density is
%preserved.

As a specific example, we will
assume that the target space is globally of the form
$K\!\times S^{N-1}$ with  the product metric
\bb
ds^2=G_{\alpha\beta}(Y)dY^\alpha dY^\beta+
V(Y)G_{ij}(\theta^i)d\theta^id\theta^j\, ,\label{metric}
\ee
where $Y^\alpha$ are coordinates on $K$. The
metric $G_{ij}$ of the coset $S^{N-1}=SO(N)/SO(N-1)$ is $Y^\alpha$
independent. One may  embed $S^{N-1}$ into $R^N$ \cite{qq}
and use
Cartesian coordinates $X^m$ instead of angular $\theta^i$. The action
is then
\bb
S[Y,X]&=&S[Y]+\int d^2z V\left(G_{mn}\p X^m\bp
X^n+\frac{1}{2\sqrt{V}R}\beta(G_{mn}X^mX^n-R^2)\right)-\nonumber \\
&&\int_{\p  \Sigma}ds A_m\frac{dX^m}{ds} \, , \label{so}
\ee
where $S[Y]=\int d^2\sigma G_{\alpha\beta}\p Y^\alpha\bp Y^\beta
-\int ds A_\beta dY^\beta/ds$ and the Lagrange multiplier $\beta$
imposes the constraint $X^2=R^2$ so that it defines $S^{N-1}$ of radius
R. Gauging this action and fixing the gauge $\Omega=\bar{\Omega}=0$ we
get the original model with target-space metric (\ref{metric}).
To find the dual action we have to fix a
gauge. A conventional such choice is to take all but one $X^m$ to
vanish, i.e.,
\bb
X^m&=&0 \, \, \, , m=1,\cdots,N-1\, ,
X^N=R\, .
\ee
Of course this choice does not completely  fix the gauge freedom since
we still have SO(N-1) invariance. We may use the remaining SO(N-1)
 gauge freedom to gauge away 1/2(N-1)(N-2) of the dual coordinates
$\La_I$.

Let us now apply the above procedure to the abelian case. Here the group is
SO(2) with the  generator $T_{mn}=\epsilon_{mn}$. The
Killing vector and the  SO(2)-invariant
U(1) gauge field are
\bb
\xi_m&=&T_{mn}X^n=\epsilon_{mn}X^n \, , \nonumber \\
A_m&=&Q(Y)\epsilon_{mn}X^n\, , \label{axi}
\ee
respectively. The corresponding  action turns out  to be
\bb
S[Y,X]&=&S[Y]+\int d^2\sigma V\left(\delta_{mn}\p X^m\bp
X^n+\frac{1}{2\sqrt{V}R}\beta(\delta_{mn}X^mX^n-R^2)\right)-\nonumber \\
&&\int_{\p  \Sigma}ds Q\epsilon_{mn}X^n\frac{dX^m}{ds} \, , \label{so1}
\ee
By gauging this action, fixing the gauge $\Omega,\bar{\Omega}=0$
 and eliminating
the Lagrange multiplier $\beta$ we get
\bb
S[Y,X]=S[Y]+\int d^2\sigma R^2V\p \theta\bp \theta-
\int_{\p  \Sigma}ds Q\epsilon_{mn}X^n\frac{dX^m}{ds} \, . \label{so2}
\ee
To obtain the dual action, we first fix the gauge $X^1=0,X^2=R$. Then,
eq.(\ref{hh}) gives
\bb
h=\partial \La\, ,&\bar{h}=-\bp \La&,\, f=VR^2\,
\ee
 and the dual action turns out to be
\bb
\tilde{S}[Y,\La]=S[Y]+\int d^2\sigma \frac{1}{VR^2}\p \La\bp \La
\, , \label{Udual}
\ee
The dual field $\La$ satisfies the boundary condition (\ref{dirr}) and since
$\Lie{\xi}A_m=\p_m\phi=0$ we have $v=QR^2 -v_0$ so that
\bb
\La+QR^2=v_0 \, \, \, on \, \, \p\Sigma \, . \label{u1}
\ee

\section{Non-abelian duals of 4D black holes}
We will apply here the formalism developed in the previous section to find
the non-abelian duals of 4D black holes.
The four-dimensional
 low-energy effective action for
open strings \cite{dorn},\cite{callann} and in the string frame is,
\bb
I_{eff}=\int d^4x\sqrt{G_\sigma}\left(e^{-2\Phi}\frac{1}{2}R
+2e^{-2\Phi}\p_\mu\Phi\p^\mu\Phi-
\frac{1}{8}e^{-\Phi}F_{\mu\nu}F^{\mu\nu}\right) \, , \label{eff}
\ee
where we have considered a U(1) gauge field only and no antisymmetric
field. The different powers
of $e^{-\Phi}$ in front of the gravitational and gauge terms appear
because the former term arises from  the sphere while the latter comes
from the disc. The field equations  in the Einstein frame
$G_{\mu\nu}=e^{-2\Phi}G_{\mu\nu}^\sigma$ are
\bb
0&=&R_{\mu\nu}-2\p_\mu \Phi\p_\nu\Phi-\frac{1}{2}e^{-\Phi}
F_{\mu\rho}{F_\nu}^\rho
+\frac{1}{8}G_{\mu\nu}e^{-\Phi}F_{\rho\lambda}F^{\rho\lambda} \, ,
\label{R} \\
0&=&\nabla^2\Phi+\frac{1}{16}e^{-\Phi}F_{\mu\nu}F^{\mu\nu}\, ,
\label{dillll}\\
0&=&\nabla_\mu(e^{-\Phi}F^{\mu\nu}) \, . \label{f}
\ee
We are looking for static, spherically symmetric solutions to the
equations above. There exist two such solutions,
a magnetically and an electrically charged black hole and we will first discuss
the former one.

 An SO(3) symmetric U(1) field strength $F_{\mu\nu}$
is the pure magnetic Maxwell field
\bb
F_m=Q\sin\th
d\th\wedge d\ph\, , \label{magn}
\ee
 where, due to the  Dirac quantization condition,
Q must be an integer multiple of 1/2.
Similarly, an SO(3)-invariant ansatz for the
four-dimensional metric is
\bb
ds^2=-\lambda^2dt^2+\frac{dr^2}{\lambda^2}+\rho^2
(d\th^2+\sin^2\th d\ph^2) \, , \label{ans}
\ee
where $\lambda,\rho$ are functions of $r$ only. Then, the solutions of
 eqs(\ref{R}--\ref{f}) are  \cite{strom},
\bb
\lambda^2&=&
\left(1-\frac{r_+}{r}\right)\left(1-\frac{r_-}{r}\right)^{3/5} \, ,
\label{l} \\
\rho&=&r\left(1-\frac{r_-}{r}\right)^{1/5} \, , \label{r} \\
e^{-\Phi_m}&=&\left(1-\frac{r_-}{r}\right)^{2/5} \, . \label{sdil}
\ee
These solutions are given in terms of two integration constants
$r_+,r_-$. The mass M and the charge Q are given in terms of these
constants as
\bb
M&=&\frac{1}{2}r_++\frac{3}{10}r_- \, , \label{Mas} \\
Q&=&4(\frac{1}{5}r_+r_-)^{1/2} \, . \label{Q}
\ee

To find an electric solution, one may use the electric-magnetic
duality $F_{mn}\to {}^*F_{mn}$ which is a symmetry of the field equations
(\ref{R}--\ref{f}). In particular, in order to convert the field equation
(\ref{f}) into the Bianchi identity, the duality rotation
\bb
{}^*F^{mn}=\frac{1}{2}e^{-\Phi}\epsilon^{mnkl}F_{kl} \, , \label{duality}
\ee has to be performed. Then, the remaining field equations are invariant if
we also transform the dilaton as $\Phi\to -\Phi$. Thus, the
electrically charged solution is
\bb
F_e&=&e^{-\Phi_m}\frac{Q}{\rho^2}dt\wedge dr=
\frac{Q}{r^2}dt\wedge dr
\, ,  \label{electric} \\
e^{\phi_e}&=&\left(1-\frac{r_-}{r}\right)^{2/5} \, , \label{ddil}
\ee
with the same metric  (\ref{ans},\ref{l},\ref{r}).

 In order to apply the duality transformation, one has to transform the
 solutions above into the string frame. In this frame, the metric is
of the form
\bb
ds^2=-\alpha^2dt^2+\frac{dr^2}{\beta^2}+\gamma^2
(d\th^2+\sin^2\th d\ph^2) \, . \label{anso}
\ee
The metric components $(\alpha,\beta,\gamma)$ are functions of $r$ only and are
explicitly given by
\bb
\alpha^2_m&=&
\left(1-\frac{r_+}{r}\right)\left(1-\frac{r_-}{r}\right)^{-1/5} \, ,
\label{l1} \\
\beta^2_m&=&
\left(1-\frac{r_+}{r}\right)\left(1-\frac{r_-}{r}\right)^{7/5} \, ,
\label{l2}\\
\gamma^2_m&=&r^2\left(1-\frac{r_-}{r}\right)^{-2/5} \, , \label{r2}
\ee
for the magnetic solution and by
\bb
\alpha^2_e&=&
\left(1-\frac{r_+}{r}\right)\left(1-\frac{r_-}{r}\right)^{7/5} \, ,
\label{l3} \\
\beta^2_e&=&
\left(1-\frac{r_+}{r}\right)\left(1-\frac{r_-}{r}\right)^{-1/5} \, ,
\label{l4}\\
\gamma^2_e&=&r^2\left(1-\frac{r_-}{r}\right)^{6/5} \, , \label{r3}
\ee
for the electric one.
The U(1) field strength is not affected by the Weyl transformation when going
from the Einstein frame to the string frame.

 The solutions given above are
 clearly SO(3) invariant and one may try to find the
dual solution. This can be done in two ways. We express the angular part of the
metric in Cartesian
coordinates $(X^1,X^2,X^3)$ and we impose the constraint
$\sum_{i=1}^3{X^i}^2=1$. We then gauge the action in the way described before
and we fix the gauge by choosing
\bb
X^1=X^2=0, \hspace{.2cm} X^3=1,\hspace{.2cm} \La_2=0\, . \label{gfixing}
\ee
In this gauge and by taking $(T_I)^m_n=\frac{1}{\sqrt{2}}\epsilon_{Inm}$ which
are properly normalized and satisfy the SO(3) algebra with
${c_{IJ}}^K=\frac{1}{\sqrt{2}}\epsilon_{IJK}$ we find by employing
eq.(\ref{hh})
\bb
h_I&=&\left( \begin{array}{ccc}
\p \La_1 & 0&\p \La_3\end{array}\right) \\
\bar{h}_I&=&\left( \begin{array}{ccc}
-\bp \La_1 & 0&-\bp \La_3\end{array}\right) \\
f_{IJ}&=&\frac{1}{\sqrt{2}} \left(\begin{array}{ccc}
\frac{1}{\sqrt{2}}\gamma^2&-\La_3&0\\
\La_3&\frac{1}{\sqrt{2}}\gamma^2&-\La_1\\
0&\La_1&0
\end{array}\right)\, .
\label{fhh}
\ee
By using  the above  expressions one may evaluate
$h_I(f^{-1})^{IJ}\bar{h}_J$ which
is found to be
\bb
h_I(f^{-1})^{IJ}\bar{h}_J=-\frac{1}{\gamma^2(x^2-y^2)}\left(\gamma^4\p y
\bp y+x^2\p x\bp x\right)\, , \label{hfh}
\ee
where, as in \cite{qq}, we have defined $x^2=2(\La_1^2+\La_3^2)$ and
$y^2=2\La_3^2$. Thus,
the dual space has a metric given by
\bb
ds^2=-\alpha^2dt^2+\frac{dr^2}{\beta^2}+\frac{1}{\gamma^2(x^2-y^2)}
\left(\gamma^4dy^2+x^2dx^2\right)\, .
\label{dualmetric}
\ee
The same result can also be found by using angular coordinates $(\th,\ph)$
for $S^2$ instead of the Cartesian $X^m$. In
these angular coordinates the Killing vectors are
\bb
\xi_1&=&\frac{1}{\sqrt{2}}(\cos \ph\p_\th-\sin \ph\cot\th\p_\ph)
 \, , \nonumber \\
\xi_2&=&\frac{1}{\sqrt{2}}(\sin\ph\p_\th+\cos\ph\cot\th\p_\ph)
 \, , \nonumber \\
\xi_3&=&\frac{1}{\sqrt{2}}\p_\ph \, . \label{killings}
\ee
We may  gauge the action in the way described in section 2,  and we fix the
gauge by choosing $\th=\pi/2,\,\ph=0,\, \La_1-\La_3=0$. After the gauge fixing,
we find that $h_I,\bar{h}_I,f_{IJ}$ are given by
\bb
h_I&=&\left( \begin{array}{ccc}
\p \La_1 & \p\La_2&-\p\La_1 \end{array}\right) \\
\bar{h}_I&=&\left( \begin{array}{ccc}
-\bp \La_1 & -\bar{\La}_2&\bp \La_1\end{array}\right) \\
f_{IJ}&=&\frac{1}{\sqrt{2}} \left(\begin{array}{ccc}
\frac{1}{\sqrt{2}}\gamma^2&-\La_1&-\La_2\\
\La_1&0&\La_1\\
\La_2&-\La_1&\frac{1}{\sqrt{2}}\gamma^2
\end{array}\right)\, .
\label{fhhh}
\ee
Proceeding as before, one may evaluate $h_I(f^{-1})^{IJ}\bar{h}_J$ and
the dual metric, as expected,  is again given by eq.(\ref{dualmetric})
with $x^2=2(2\La_1^2+\La_2^2)$ and $y^2=2\La_2^2$
 this time.

Let us now turn to the boundary conditions. In the electric case, there
is no U(1) field in the $(\th,\ph)$ direction and
 the dual coordinates $(x,y)$ are
restricted to satisfy the Dirichlet conditions
\bb
x=x_0\, , \,\,\, y=y_0 \, \, \, on\, \, \p\Sigma\, , \label{bps}
\ee
where $(x_0,y_0)$ are constants.
For the magnetic case, the gauge
potential in the upper hemisphere is $A=Q(1-\cos\th)d\ph$. Then,  by employing
eq.(\ref{Liee}) we find
\bb
\phi_1&=&\frac{1}{\sqrt{2}}\sin \ph\left(\frac{1}{\sin\th}-\cot\th\right)\,
,\nonumber \\
\phi_2&=&-\frac{1}{\sqrt{2}}\cos \ph\left(\frac{1}{\sin\th}-\cot\th\right)\,
,\nonumber \\
\phi_3&=&\phi_3^0 \, , \label{sdf}
\ee
and from eq.(\ref{uh}) we get
\bb
v_1&=&\frac{Q}{\sqrt{2}}\frac{\sin\ph}{\sin\th}+v_1^0 \, , \nonumber \\
v_2&=&\frac{Q}{\sqrt{2}}\sin\ph\sin\th +v_2^0\, , \nonumber \\
v_3&=&\frac{Q}{\sqrt{2}}(1-\cos\th)+v_3^0\, , \label{uuu}
\ee
where $(v_1^0,v_2^0,v_3^0)$ are constants. After gauge fixing the dual
fields turn out to satisfy the Dirichlet conditions (\ref{bps}).
It should be noted that after the gauge fixing, there remain two independent
parameters out of $(v_1^0,v_2^0,v_3^0)$ which represents the freedom in the
position of the D-brane.

\section{Conclusions}

We have examined here the non-abelian duality transformations for
open strings following  the path integral approach \cite{rv}.
There exist some  peculiarities due to the
boundary \cite{Oalv}.   Dirichlet boundary conditions
in the dual model arise as in the abelian case.
 This shows that also for open strings moving in a background
with non-abelian isometries D-branes appear in a natural way. It should be
stressed that the Dirichlet conditions seem to be imposed in the dual model as
external conditions irrespectively of any stationary condition of the
action. This is because one integrates the gauge fields on the boundary
independently of the gauge fields on the bulk \cite{otto} and it is this
integration that  produces the Dirichlet
conditions.

We have applied the non-abelian duality transformation in the case of 4D
target spaces. In particular we have discussed an electrically and a
magnetically charged black hole. For both of these solutions we have presented
their non-abelian duals. The dual coordinates satisfy
 Dirichlet conditions which define curved 1-branes. The world volume of these
 branes has black-hole structure as well.
The singularity is at $r=r_-$, the horizon
is at $r=r_+$ and for the extremal case $r_+=r_-$ we have a naked singularity.
Thus, the dual model contains  extended objects, namely  D-branes, which
will act as a source for the closed string modes, the graviton and the dilaton
\cite{Duff}.
Conformal invariance is then imposed by the Born-Infeld
action for the D-brane \cite{dai}
which, moreover, will also specify the dilaton shift.
In our case,  the dual electric and magnetic
solutions describe
fundamental and  solitonic 1-branes, respectively, since in the former case we
expect a source term due to a non-vanishing electric field,
 while such a term is absent in the latter case since
the coupling to the  U(1) gauge field vanishes due to Dirichlet conditions.
A detailed discussion of these issues including
 the dilaton is under investigation.

 In our discussion we did not include global issues. For
non-abelian dualities this is an outstanding problem for the case of closed
strings, too.
A further topic to be addressed in a future work is the extension of
non-abelian T-duality to the open superstring as well as the inclusion of
Chan-Paton factors.

\vspace{.5cm}

\noindent
{\bf Acknowledgement}\\
We would like to thank H. Dorn for discussion and for
providing us with an early
version of  \cite{otto}. The work of S.F. is supported by GIF, German-Israeli
Foundation for Scientific Research. A.K. is supported by the Alexander von
Humboldt-Stiftung. The work of S.S. is supported by DFG, Deutsche
Forschungsgemeinschaft.

\end{document}